\pdfoutput=1
\documentclass[12pt,a4paper]{article}
\usepackage[utf8]{inputenc}
\usepackage{a4wide}
\usepackage{amsmath,amssymb,amsfonts}
\usepackage{subfig,graphicx}
\usepackage[dvipsnames]{xcolor}
\usepackage[colorlinks,allcolors=blue]{hyperref}
\usepackage{dsfont}

\newcommand{\beq}{\begin{eqnarray}}
\newcommand{\eeq}{\end{eqnarray}}
\newcommand{\non}{\nonumber\\}
\DeclareMathOperator{\U}{U}
\DeclareMathOperator{\SU}{SU}

\newcommand{\p}{\partial}
\renewcommand{\i}{\mathrm{i}}
\renewcommand{\d}{\mathop{}\!\mathrm{d}}

\newcommand{\dH}{\delta{\mkern-2.5mu}H}
\newcommand{\dL}{\delta{\mkern-1.5mu}L}
\newcommand{\dS}{\delta{\mkern-1.5mu}S}

\DeclareMathOperator{\tr}{tr}
\DeclareMathOperator{\Tr}{Tr}

\DeclareMathOperator{\MeV}{MeV}
\DeclareMathOperator{\fm}{fm}
\DeclareMathOperator{\km}{km}
\newcommand{\calA}{\mathcal{A}}
\newcommand{\calE}{\mathcal{E}}
\newcommand{\calF}{\mathcal{F}}

\newcommand{\hA}{\widehat{A}}

\newcommand{\hF}{\widehat{F}}

\newcommand{\wha}{\widehat{a}}
\newcommand{\bchi}{\boldsymbol{\chi}}
\newcommand{\btau}{\boldsymbol{\tau}}

\begin{document}
\pagenumbering{Roman}
\begin{titlepage}
  \begin{flushright}
    February 2024
  \end{flushright}
  \vskip3cm
  \begin{center}
    {\Large\bf Boundary terms in the Witten-Sakai-Sugimoto model at
      finite density}\\[2cm]
    {\bf Lorenzo Bartolini, Sven Bjarke Gudnason}\\[2cm]
    {Institute of Contemporary Mathematics, School of Mathematics and Statistics,\\ Henan University, Kaifeng, Henan 475004, P. R. China}
  \end{center}
  \vfill
  \begin{abstract}
    We consider the Witten-Sakai-Sugimoto model in the approximation
    of smeared instantons at finite density via a homogeneous Ansatz,
    which is known to be discontinuous in order to be able to contain
    a nonvanishing baryon density.
    The discontinuity at the infrared tip of the bulk spacetime gives
    rise to subtleties of discarding boundary terms that are normally
    discarded in the literature.
    We propose a reason for discarding this boundary term, by
    scrutinizing the currents and topological properties of the model.
    Along the way, we find a very effective and simple condition to
    compute the point of thermodynamic equilibrium.
  \end{abstract}
  \vfill
  \rule{10cm}{0.5pt}\\
  {\tt lorenzo@henu.edu.cn}, {\tt gudnason@henu.edu.cn}
\end{titlepage}

\pagenumbering{arabic}

\tableofcontents

\section{Introduction}

Chern-Simons (CS) terms exist in field theories of odd (spacetime)
dimensions, most famously perhaps in three dimensions.
The CS theory by itself seems at first glance to be quite
uninteresting, since it is a topological theory and possesses no
dynamics.
This property changes, however, once the theory is coupled to
Yang-Mills (YM) (or Maxwell) theory \cite{Deser:1981wh,Deser:1982vy}
or even just to a scalar field theory \cite{Jackiw:1990aw}.
The coupling of CS theory induces immediately new behavior into the
theory with which one couples it to; for example, the traditional
Gauss law is modified and if CS and YM theories are both present, the
gauge field propagator will have a richer pole structure, yielding a
topologically massive theory \cite{Deser:1981wh,Deser:1982vy}. 

Another aspect of CS terms is that they are not manifestly gauge
invariant, unlike the YM ($\tr F\wedge*F$) or Maxwell ($F$ and hence $F\wedge*F$)
counterparts, where $F$ is $\mathfrak{g}$-valued with $\mathfrak{g}$
being the algebra corresponding to the Lie group $G$ (for Maxwell the
generator of $\U(1)$ is just a real number).
This is immediately clear from the fact that the CS term contains the
gauge field $A$ in addition to the field strength $F=\d A+\i A\wedge A$,
and since $A$ transforms like $A\to A+\d_A\eta$ under an infinitesimal
gauge transform $\eta$, where $\d_A\eta=\d\eta+\i[A,\eta]$ is the gauge
covariant derivative and $\eta$ is also $\mathfrak{g}$-valued.
The field strength transforms covariantly and hence the
trace of any power of the field strength is invariant under gauge
transformations. In particular, YM theory is manifestly
gauge invariant.
The CS term on the other hand transforms into (itself plus) a total
derivative and plus a winding number of the gauge fields.
For suitably chosen integer coefficients (when the CS term is
appropriately normalized), the latter winding number term yields a
contribution of $2\pi k$ to the action, under which $e^{\i S}$ does
not change.
The total derivative term is normally not causing any trouble in field
theories for two reasons: Physicists often work on infinite Cartesian
spaces like $\mathbb{R}^3$ or $\mathbb{R}^5$ and the fields are almost
always assumed to be continuous and differentiable.

A counterexample to the first reason, leads to beautiful results in
condensed matter theory when the CS term is utilized for the
fractional quantum Hall effect \cite{Zhang:1988wy} and there exists
a boundary, where the fields are not necessarily pure gauge, the
boundary effects give rise to a phenomenon of edge modes living on
the boundary (circle) of the material \cite{Wen:1990se}. 

A counterexample to the latter reason, on the other hand is what we
are concerned with in this paper. We are interested in holographic
nuclear matter, which is the situation in which we have large
densities of matter in holographic QCD \cite{Kim:2012ey}. To be
specific, we are considering the popular top-down holographic
QCD model, namely that of Witten-Sakai-Sugimoto (WSS)
\cite{Witten:1998zw,Sakai:2004cn}\footnote{
  See also Refs.~\cite{Hong:2007kx,Hong:2007ay} for a derivation of
  the chiral effective action from the WSS model, as well as an
  estimate of the axial coupling, magnetic dipole moments,
  electromagnetic form factors and vector dominance.
}.
This model at low energies is indeed described by 5-dimensional
YM and CS terms coupled together in a curved anti-de
Sitter-like spacetime.
This particular theory has a conformal boundary with a finite
curvature, which is known as the UV boundary where the UV degrees of
freedom live.
The holographic principle states that the theory in the bulk is dual
to the field theory living on the conformal boundary and observables
of the bulk fields can be read off of the tails of the fields near the
conformal boundary.
This should already raise concern for the astute reader, since
nontrivial or non-pure gauge behavior on a boundary could spell
trouble for the theory.
It turns out that this gauge variance is welcome, as it reproduces the
chiral anomaly of QCD \cite{Sakai:2004cn}.\footnote{There are subtleties for
gauge invariance of the CS term when topologically nontrivial gauge
configurations are considered and when the gauge group is $\SU(N)$
with $N>2$, see Ref.~\cite{Lau:2016dxk}. }
The above-mentioned second issue with CS, is when the fields are not
continuous.
This issue arises only in the limit of finite density baryonic matter
in the approximation of homogeneous matter in the bulk, where it has
been proved that no such continuous configurations can exist
\cite{Rozali:2007rx}.
It is possible, however, to describe homogeneous nuclear matter at
finite/large densities if we allow the gauge field configurations to
be discontinuous at the IR tip of the cigar-shaped spacetime of the
WSS model \cite{Li:2015uea}.\footnote{Instead of employing a
discontinuity in the fields, it is possible to impose asymmetric
boundary conditions on the fields \cite{Elliot-Ripley:2016uwb}. }
This would correspond to a smeared configuration of baryons/instantons.

This is where we meet the issue with the CS term, in particular
because the most convenient form of the CS term for use in the
baryonic sector, as written down in Ref.~\cite{Hata:2007mb} as the Abelian
``electric'' field multiplied by $\tr F\wedge F$. This formulation of
the 5-dimensional CS term turns out to be natural for homogeneous
nuclear matter, but differs with the full CS term $\omega_5(A)$ by a
boundary term.
In this paper, we explore the difference of the WSS model when taking
this boundary term into account or discarding it.
It turns out that the holographic dictionary and the thermodynamical
laws are well defined in either case, but that there is a preferred
choice if we scrutinize the currents of the theory.
In particular, matching the baryon charge and the behavior of the
fields at the conformal boundary provides a way to choose which
boundary terms to discard.
We additionally find that the same choice of the CS term, makes this
term invariant under the $\SU(2)$ gauge transformation that is needed
to show the equivalence between isospin realized by isorotations of
the fields and turning on a chemical potential at infinity.

This paper is organized as follows.
Sec.~\ref{sec:CS_from_ST} reviews the derivation of the CS term from
the point of view of the WSS model in string theory.
In Sec.~\ref{sec:homogeneous}, we set up the notation for the
homogeneous Ansatz for nuclear matter in the model at hand.
In Sec.~\ref{sec:thermo}, we present a systematic way to derive the
thermodynamic equilibrium conditions, which turns out to be very
useful for numerics.
In Sec.~\ref{sec:energymomentum}, we show via the energy momentum
tensor that the standard thermodynamic relations work, regardless of
whether the boundary term in CS is included or not.
In Sec.~\ref{sec:observables}, we illustrate the difference between
taking the boundary term in the CS term into account or not, by
computing a range of observables.
We conclude the paper in Sec.~\ref{sec:conclusion} with a discussion.
Details of the equivalence between $\SU(2)$-isospin rotation of the
fields and the introduction of an external chemical potential at
infinity are shown in App.~\ref{app:chempot}, whereas the details on
how the chiral anomaly of QCD is unchanged by our proposal are
delegated to App.~\ref{app:chiralanomaly}.

\section{Chern-Simons term from string theory}\label{sec:CS_from_ST}

D$p$-Branes are described by the DBI action:
\beq
S_{\rm DBI} = -T_p\int\d^{p+1}\xi e^{-\Phi}\sqrt{-\det\left(\gamma_{ab}+B_{ab}+2\pi\alpha'F_{ab}\right)},
\eeq
where $F,B,\gamma$ are respectively the gauge field strength, the
Kalb-Ramond form and the induced metric on the $(p+1)$-dimensional world
volume. The overall brane tension $T_p$ can be given terms of in
string parameters as $T_p=(2\pi)^{-p}\alpha'^{-\frac{p+1}{2}}$. 
We will be interested in the $B_{ab}=0$ scenario, and we will employ
the Yang-Mills approximation of this action, obtained by expanding the
square root and keeping the quadratic order in the field strength. 
Following Refs.~\cite{Sakai:2004cn,Sakai:2005yt,Hata:2007mb}, we write
the resulting Yang-Mills action in the form 
\beq
S_{\rm YM}=-\kappa \Tr\int\d^4x\d z\left[\frac{1}{2}h(z)\calF_{\mu\nu}^2+k(z)\calF_{\mu z}^2\right],\qquad
\kappa=\frac{\lambda N_c}{216\pi^3}.
\eeq
The other coupling present in the case of a stack of D$p$-Branes is
given by world volume coupling to the Ramond-Ramond (RR) forms. As
argued in Ref.~\cite{Green:1996dd}, in the presence of a D-Brane
background inducing a nontrivial flux of an RR form, the correct 
expression of this coupling to be considered, is the one after
integration by parts, where the RR field strength appears
explicitly. Since we will work in a setup with only the flux of
$F_4$ turned on, we take the coupling to be 
\beq
S_{\rm CS}=\frac{1}{48\pi^3}\int_{D8}F_4\wedge\omega_5(\calA),\qquad
\omega_5(\calA)
=\Tr\left(\calA\wedge\calF^2-\frac{\i}{2}\calA^3\wedge\calF-\frac{1}{10}\calA^5\right),
\eeq
where powers of forms are understood with the wedge product.
Assuming now dependence of the gauge fields only on coordinates
transverse to $S^4$, we can integrate out $F_4$ using its flux to
obtain 
\beq
S_{\rm CS}=\frac{N_c}{24\pi^2}\int_{5D}\omega_5(\calA).
\eeq
In the general case, we will have an arbitrary number $N_f$ of flavor
branes (although for $N_f>N_c$ it would be appropriate to include the
backreaction of the branes onto the geometry), hence it is possible to
write $S_{\rm CS}$ separating out the $\SU(N_f)$ part:
\beq
S_{\rm CS}=\frac{N_c}{24\pi^2}\int_{5D}\Tr\left[\omega_5^{\SU(N_f)}+3\widehat{A}\wedge F^2+\widehat{A}\wedge\widehat{F}^2+\d\left(\widehat{A}\wedge\left(2F\wedge A-\frac{\i}{2}A^3\right)\right)\right],
\eeq
where $\mathcal{F} = F + \widehat{F}$ splits the field strength into
the non-Abelian and Abelian part, respectively, and similarly for the
gauge field $\mathcal{A}$.
Since we consider the $N_f=2$ case, accounting only for the existence
of two light flavors (of quarks), the first term proportional to
$\omega_5^{\SU(2)}$ vanishes, leaving us with the simpler expression: 
\beq\label{SCSfull}
S_{\rm CS}=\frac{N_c}{24\pi^2}\int_{5D}\Tr\left[3\widehat{A}\wedge F^2+\widehat{A}\wedge\widehat{F}^2+\d\left(\widehat{A}\wedge\left(2F\wedge A-\frac{\i}{2}A^3\right)\right)\right].
\eeq
If we assume the $\SU(2)$ fields, $A^a$, to be continuous functions
vanishing at spatial infinity fast enough for the configuration to
have a finite energy, then the total-derivative term vanishes and we
are left with the commonly used expression of Chern-Simons (CS) action in
the Witten-Sakai-Sugimoto model. If we employ the field expansion 
\begin{align}
A&={A}_{\alpha }^a T^a \d x^\alpha,\qquad
\widehat{A}=\widehat{A}_{\alpha}\frac{\mathds{1}}{2}\d x^\alpha, \non
\alpha,\beta,\ldots &= {0,M},\qquad M,N,\ldots = {i,z},\qquad i,j,\ldots = {1,2,3},
\end{align}
then the resulting action term reads
\beq
\label{SCSbulk}
S_{\rm CS}=\frac{N_c}{384\pi^2}\epsilon^{\alpha_1\alpha_2\alpha_3\alpha_4\alpha_5}\int\d^4x\d z
\hA_{\alpha_1}\left[3F^a_{\alpha_2\alpha_3}F^a_{\alpha_4\alpha_5} +\hF_{\alpha_2\alpha_3}\hF_{\alpha_4\alpha_5}\right].
\eeq
The main focus of this article is to study a particular situation, the
homogeneous Ansatz, in which the assumptions for reducing the
CS term as above, are not justified, leading to a
nonvanishing contribution from the total-derivative term in
Eq.~\eqref{SCSfull}.

\section{The homogeneous Ansatz}\label{sec:homogeneous}

In holographic QCD, baryons are realized as topological solitons of
the bulk theory, whose holonomy produces Skyrmions of the boundary
theory \cite{Atiyah:1989dq}.
A system of baryons like a nucleus or in general nuclear
matter is then realized as a many-soliton configuration in a
five-dimensional curved spacetime: This problem is in general very
difficult to solve even using numerical methods, and a variety of
approximations are usually employed to make it more treatable. Note
that topological solitons in this model are indeed smooth
configurations of the gauge fields, with the exception of a
singularity at $\xi=0$, with $\xi^2\equiv (\vec{x}-\vec{X})^2+(z-Z)^2$, which is
however a gauge artifact.
Hence, dealing with this description of baryons we can safely drop the
total derivative term in the CS action.

The homogeneous Ansatz is an approximation employed in many
holographic setups to describe matter at high density.
We employ the approximation of forming a homogeneous fluid, where
the single baryons are no longer identifiable, smeared along the three
dimensions of space of the boundary theory.  
To realize this, we assume the fields to depend only on $z$, and in
the static scenario only the following fields are turned on (see
Ref.~\cite{Bartolini:2022gdf} for a generalization including time
dependence in the form of a slow $\SU(2)$
rotation):\footnote{The homogeneous Ansatz has also been
  employed in bottom-up holographic QCD models, such as VQCD
  \cite{Ishii:2019gta,Jarvinen:2021jbd} and the hard-wall model
  \cite{Bartolini:2022rkl}.}
\beq
A_i=-\frac{H(z)}{2}\tau^i,\qquad
\hA_0=\widehat{a}_0(z),
\label{eq:hom_ansatz}
\eeq
where we use the $A_z=0$ gauge as done in Ref.~\cite{Li:2015uea}\footnote{A general gauge
    transformation consistent with the requirement of homogeneity
    $A_M\to U(z) (A_M - \i\p_M) U(z)^{-1}$ alters only $A_z$ whereas
    the field strengths transform covariantly. The topological charge
    and the YM action remain invariant, but so does the Chern-Simon
    action since $F_{i0}$ vanishes in the static homogeneous Ansatz.}.
The complicated many-soliton problem in four-dimensional space is now
substituted by the simpler physics of continuous matter in one
dimension. The price we pay for this great simplification is the loss
of any information on the properties of the individual baryons, and their
configuration in space (e.g.~we lose information about the favored
lattice configuration at a certain density).
In particular, we compute the baryon number, which will be infinite
given the homogeneity of the system, but most of all will not be
quantized in integers. The only meaningful quantity within this 
Ansatz is then the baryon density: Usually this quantity is encoded
in the field strengths, $F_{MN}$, whereas now it will be encoded in the
function $H(z)$ as 
\begin{align}
d&=\frac{1}{32\pi^2}\int\d z \epsilon^{MNPQ}\Tr F_{MN}F_{PQ}\non
&=-\frac{1}{8\pi^2}\int\d z \partial_z\left(H^3\right)\non
&=-\frac{1}{8\pi^2}\left[H^3\right]^{z=+\infty}_{z=0^+}-\frac{1}{8\pi^2}\left[H^3\right]_{z=-\infty}^{z=0^-}.
\end{align}
For the \emph{energy density} to be finite, the function $H(z)$ has to vanish at
$z\to\pm\infty$, so that we remain with
\beq
d=\frac{1}{8\pi^2}\left[H^3(z\rightarrow0^+)-H^3(z\rightarrow0^-)\right].
\eeq
We see that the only way to have a finite density, and so to describe
nuclear matter with this Ansatz, is for the function $H(z)$ to have a
discontinuity: We choose the value $z=0$ for the location of this
discontinuity because it is the energetically favored position at low
densities, as can be inferred by the semiclassical value of the
pseudomodulus $Z_{cl}=0$ of the single baryon configuration
\cite{Hata:2007mb}, but in principle the location of the discontinuity
should be determined by minimization of the free energy.

Requiring $H(z)$ to be odd, thus leading to a continuous field
strength, we finally obtain that the density is given by its infrared
boundary condition as 
\beq
H(z\rightarrow0^{\pm}) = \pm\left(4\pi^2 d\right)^{\frac{1}{3}}.
\eeq
From this result, we are brought to the conclusion that we cannot drop
the total-derivative term in Eq.~\eqref{SCSfull}, which will instead in
principle contribute to the energy and free energy of the system. 
Before moving to the discussion regarding the physical effects of this
term, we write the full action of the flavor fields when the static
homogeneous Ansatz is employed:
\begin{align}
S&=S_{\rm YM} + S_{\rm CS}^{\rm bulk} + S_{\rm CS}^\partial ,\\
S_{\rm YM}&=-\kappa\int \d^4x\int_0^{\infty}\d z\left[3hH^4+3k(H')^2-k(\widehat{a}_0')^2
  \right],\label{SYM}\\
S_{\rm CS}^{\rm bulk}&= -\frac{3 N_c}{8\pi^2}\int \d^4x\int_0^{\infty} \d z\;\widehat{a}_0 H'H^2 ,\label{eq:CSbulk}\\
S_{\rm CS}^{\partial}&= \frac{3N_c}{32\pi^2}\int \d^4x \int_{0}^{\infty}\d z\;\partial_z\left(\widehat{a}_0 H^3\right)
=-\frac{3N_c}{32\pi^2}\int \d^4x\;\widehat{a}_0(0) H^3(0) ,
\end{align}
where we have turned every integration over $z\in[-\infty,\infty]$
into one over $z\in[0,\infty]$ with a factor of two coming from the
other halves of the branes. Adopting this "folding", on top of making
it more convenient to deal with integrations -- since we can avoid
handling discontinuous functions, also makes the necessity of
$S_{\rm CS}^\partial$ more manifest, since we effectively have a
``boundary'' at $z=0$ which will give a nonvanishing contribution
(while the UV contribution will still vanish due to the boundary
condition $H(\infty)=0$). 
The equations of motion derived from this action are:
\beq
&&\p_z\left(k(z)H'\right)-2h(z)H^3+\frac{N_c}{16\pi^2 \kappa}H^2 \widehat{a}_0'=0,\label{eqH}\\
&&\p_z\left(k(z) \widehat{a}_0'\right)+ \frac{3N_c}{16\pi^2 \kappa}H^2H'=0 \label{eqa0}.
\eeq
Note that $S_{\rm CS}^{\partial}$ being a boundary term, does not
contribute to the equations of motion, but as we will see, it enters
physics via a different mechanism.

\section{Thermodynamic equilibrium}\label{sec:thermo}

Solving the equations of motion by itself does not guarantee that the
system is in its equilibrium configuration. Boundary conditions encode
physical quantities: Quark chemical potential $\mu$ and baryon density
$d$ are introduced as 
\beq
\widehat{a}_0(\infty) = \mu,\qquad
H(0) = \left(4\pi^2 d\right)^{\frac{1}{3}}.
\eeq
We are left with two more boundary conditions to impose: For the field
$H(z)$ we require $H(\infty)=0$ to have finite energy density, while
for the field $\widehat{a}_0(z)$ usually a Neumann condition is
employed at $z=0$ since it is an even and continuous function. However
this choice is not always justified: We will show that the new term
introduced in the CS action can change this prescription.
We start by requiring our field configuration to extremize the action:
This procedure includes imposing the equations of motion, but is not
limited to it, as boundary term arise when we integrate by parts to
make the equations of motion manifest. These boundary terms are in
many cases vanishing when requiring Dirichlet or Neumann conditions
for the fields, but our new effective boundary at $z=0$ introduces a
non-triviality in this process.

When we vary the fields as $H\rightarrow H+\dH$ and
$\widehat{a}_0\rightarrow\widehat{a}_0+\delta\widehat{a}_0$, the
action is varied as $S\rightarrow S=\dS$ with 
\beq
\dS = \int \d^4x\d z\left[\left(\text{E.o.M.}\right)_{H}\dH + \left(\text{E.o.M.}\right)_{\widehat{a}_0}\delta\widehat{a}_0\right]+\dS_{\rm boundary},
\eeq
where the first term (in the bracket) vanishes upon imposition of the
equations of motion. The explicit form of the second term is for now omitted
as we will review first the case in which we neglect the presence of
$S_{\rm CS}^\partial$, to then move to the analysis of the full
CS.

\subsection{Without \texorpdfstring{$S_{\rm CS}^\partial$}{S\_CS**del}}

If we assume that the part of the CS term,
$\dS_{\rm CS}^\partial$, vanishes then the boundary term in the
variation of the action is given by
\beq
\dS_{\rm boundary} = 2\kappa \int \d^4x \left[k(z)\widehat{a}_0' \delta\widehat{a}_0-3\left(k(z)H'+\frac{N_c}{16\pi^2\kappa}\widehat{a}_0 H^2\right)\dH\right]^{z=\infty}_{z=0}.
\eeq
We now proceed, as with the equations of motion terms, to require
these two terms (proportional to $\delta\widehat{a}_0$ and $\dH$) to
vanish separately. As a first simplification we note that the 
contribution at $z=\infty$ vanishes: In fact
$\delta\widehat{a}_0(\infty)=0$ because we work at a fixed
$\widehat{a}_0(\infty)=\mu$, and $\dH(\infty)=0$ for the energy density
to be finite. We are thus left with two additional equations to
satisfy on top of the equations of motion to truly extremize the
action: 
\begin{align}
\widehat{a}_0'(0)\delta \widehat{a}_0(0)&=0,\label{bdyeqa0OLD}\\
\left(H'(0)+\frac{N_c}{16\pi^2\kappa}\widehat{a}_0(0) H^2(0)\right)\dH(0)&=0\label{bdyeqHOLD}.
\end{align}
To solve Eq.~\eqref{bdyeqa0OLD} we can simply enforce Neumann boundary
conditions $\widehat{a}_0'(0)=0$, consistently with common
practice.
At first sight it would also seem that Eq.~\eqref{bdyeqHOLD}
is satisfied with our choice $H(0)=\left(4\pi^2d\right)^{\frac{1}{3}}$
which seems to imply $\dH(0)=0$. However, we have to keep in mind that
we are not working at fixed density, but rather at fixed chemical
potential, hence $d(\mu)$ is a dynamical quantity that should be
determined from the minimization of the action itself. This leads us
to enforce the condition
\beq\label{eqEQOLD}
H'(0)+\frac{N_c}{16\pi^2\kappa}\widehat{a}_0(0) H^2(0)=0.
\eeq
We recognize the above equation as the condition that was only
numerically verified in Appendix D of Ref.~\cite{Bartolini:2022rkl} as part
of (D.16): We now see that
indeed the condition is exactly satisfied upon extremization of the
action, which corresponds to the equilibrium configuration, at least for
every $\mu\geq\mu_{\rm onset}$.\footnote{In Ref.~\cite{Bartolini:2022rkl} the
relation is obtained for the hard-wall model. It can be mapped to
relation \eqref{eqEQOLD} in the Witten-Sakai-Sugimoto via
$M_5\rightarrow\kappa$, $h(z)\rightarrow a(z)$, $k(z)\rightarrow a(z)$ and a
change in sign due to different conventions adopted and the different
orientation of the integration domain. All considerations we make for
the Witten-Sakai-Sugimoto model translate directly to the hard-wall
model, where however the formalities concerning the choice of the
CS term are less stringent given the bottom-up nature of the
model.}
Since the field $\widehat{a}_0$ carries information on $\mu$,
while $H$ carries information about the density, we can see this
equation as giving us the relation between these two quantities at
equilibrium. It is easier to see it as defining $\mu(d)$: To make it
explicit, we write $\widehat{a}_0$ as:
\beq
\widehat{a}_0(z)= \mu - \int_{z}^{\infty} \d z'\;\widehat{a}_0'(z'),\qquad
\widehat{a}_0'= -\frac{N_c}{16\pi^2\kappa}\frac{1}{k(z)}\left(H^3(z)-H^3(0)\right)\label{da0old},
\eeq
where the second relation comes from the integration of the equation
of motion. 
After rearranging to isolate $\mu$ (which does not enter the equations
of motion, so it does not affect the values of $\widehat{a}_0'(0)$ or
$H'(0)$), we find 
\beq
\mu(d) =
-\frac{4\kappa}{N_c}\left(\frac{4\pi^2}{d^2}\right)^{\frac{1}{3}} H'(0)
+\frac{N_c d}{4\kappa}\int_0^\infty \d z \frac{1}{k(z)}\left(1-\frac{H^3(z)}{H^3(0)}\right),\label{eq:mud_bulk}
\eeq
which allows us to compute $\mu(d)$ for every given field configuration
satisfying the equations of motion. In the baryonic phase
($\mu\geq\mu_{\rm onset}$) the relation is invertible (the inversion
has to be performed numerically), giving us a simple way to compute
$d(\mu)$.
What we found is that requiring not only the equations of motion to be
satisfied, but the more general extremization of the action including
nontrivial boundary terms, we obtain the thermodynamic equilibrium
condition on top of the field configuration. Note that we have
no information about what phase is favored, so the equilibrium found
this way may very well be unstable: The baryonic phase will be the stable
only when $\Omega_{B}<0$, with $\Omega_B$ being the grand canonical
potential in the baryonic phase
(and we made use of the trivial $\Omega_{\rm vacuum}=0$).

One more detail we need to pay attention to, is the identification of
the parameters we choose to describe the chemical potential, the
baryonic density, and the physical quantities: From the holographic
dictionary we can read off the physical baryon density $d_B$ from the
expansion of $\widehat{a}_0$ near the boundary. This task is
straightforward since we have an explicit expression for
$\widehat{a}_0'(z)$, so that we can directly use the formula derived
in Ref.~\cite{Hashimoto:2008zw}: 
\beq\label{dbold}
d_B = \frac{2}{N_c}\kappa\left[k(z)\widehat{a}_0'\right]^{\infty}_{-\infty}
= \frac{4}{N_c}\kappa\left[k(z)\widehat{a}_0'\right]_{z=\infty}
= d,
\eeq
so in this case the parameter $d$, introduced as a boundary condition,
really coincides with the physical baryon density.
With this result, we now extract the physical expression of the baryon
chemical potential $\mu_B$ in terms of $\mu$: We do so by looking at
the action terms linear in $\mu$. Since $\mu$ appears only as an
overall shift of $\widehat{a}_0$, only the CS action
contains such a term, which reads 
\beq
\mu_B d_B
= -\frac{N_c}{8\pi^2}\mu \int_0^\infty\d z\;\partial_z(H^3)
= \frac{N_c}{2}\mu d,
\eeq
from which follows the identification
\beq\label{mubold}
\mu_B= \frac{N_c}{2}\mu.
\eeq

\subsection{Including \texorpdfstring{$S_{\rm CS}^\partial$}{S\_CS**del}}

We now wish to include the presence of $S_{\rm CS}^{\partial}$ in our
considerations. We again wish to extremize the action: The equations
of motion (unchanged from the previous section) will take care of the
bulk contribution, leaving us again with the need to satisfy a pair of
equations on the IR boundary. The novelty with respect to the previous
section is that now $\dS_{\rm CS}^{\partial}$ will provide additional
boundary terms, modifying Eqs.~\eqref{bdyeqa0OLD} and \eqref{bdyeqHOLD}:
\begin{align}
\left(\widehat{a}_0'(0) +  \frac{3 N_c}{64\pi^2 \kappa} H^3(0) \right)\delta \widehat{a}_0(0)&=0,\label{bdyeqa0NEW}\\
\left(H'(0)+\frac{N_c}{16\pi^2\kappa}\widehat{a}_0(0) H^2(0) - \frac{3 N_c}{64\pi^2\kappa}\widehat{a}_0(0) H^2(0)\right)\dH (0)&=0\label{bdyeqHNEW}.
\end{align}
Both equations acquire a new term, but most notably the new boundary
equation for $\widehat{a}_0$ is no longer satisfied by a Neumann
boundary condition, which has to be modified to
\beq
\widehat{a}_0'(0) = -\frac{3N_c}{64\pi^2 \kappa} H^3(0),
\eeq
which in turn leads to a different solution to the integrated equation
of motion \eqref{eqa0}: 
\beq\label{da0new}
\widehat{a}_0'=-\frac{1}{k(z)}\frac{N_c}{16\pi^2\kappa}\left( H^3(z)-\frac{H^3(0)}{4}\right).
\eeq
As a consequence, also the thermodynamic equilibrium relation $\mu(d)$
gets modified, both by the new term in Eq.~\eqref{bdyeqHNEW} and the
new solution \eqref{da0new}: 
\beq
\mu(d) =-\frac{16\kappa}{N_c}\left(\frac{4\pi^2}{d^2}\right)^{\frac{1}{3}} H'(0)
+\frac{N_c d}{4\kappa}\int_0^\infty \d z \frac{1}{k(z)}\left(\frac{1}{4}-\frac{H^3(z)}{H^3(0)}\right).\label{eq:mud_full}
\eeq

Repeating the argument from the previous section, we now want to map
the parameters $d,\mu$ to the physical quantities $\mu_B,d_B$: Again
we can compute the baryon density from the asymptotics of the
$\widehat{a}_0$ field according to the formula for the
current\footnote{A naive approach to the calculation of this charge is
to trade the evaluation at the boundary for the integral over $z$ of
the derivative of the expression, and then use Eq.~\eqref{eqa0}
to obtain exactly the instanton number density $d$. However, since
$k\wha_0'$ is not continuous in this scenario, we cannot exchange the
function evaluated at the UV boundary for the bulk integral of
the derivative, as it would pick up IR contributions that do not
belong to the definition of the current. }, which 
however now yields a different result because of the new expression
\eqref{da0new}:
\beq\label{dbnew}
d_B = \frac{2}{N_c}\kappa \left[k(z)\widehat{a}_0'\right]^{\infty}_{-\infty}
= \frac{4}{N_c}\kappa\left[k(z)\widehat{a}_0'\right]_{z=\infty}
= \frac{d}{4}.
\eeq
Since the asymptotic leading order for $\widehat{a}_0$ is unchanged,
we expect $\mu_B$ to be unchanged, and this is confirmed by looking at
the action terms proportional to $\mu$, now coming both from
$S_{\rm CS}^{\rm bulk}$ and $S_{\rm CS}^\partial$, amounting to a contribution:
\beq
S_{\mu} = -\frac{N_c}{8}\mu d = -\frac{N_c}{2}\mu d_B
\eeq
hence we can still rely on the identification \eqref{mubold}.

\section{Free energy, energy and pressure}\label{sec:energymomentum}

In Appendix D of Ref.~\cite{Bartolini:2022rkl} the computation of the
energy density and pressure from the stress-energy tensor of the
flavor fields was presented: The computation illustrated a number of
non-trivialities, including contributions of boundary terms for both
quantities, and the use of the newly formulated boundary equations
\eqref{bdyeqHOLD} and \eqref{bdyeqHNEW}.
Here, we repeat the calculation in both the cases with
$S_{\rm CS}^\partial =0$ and with $S_{\rm CS}^\partial \neq 0$,
showing how everything comes consistently together and keeping the
holographic dictionary entry $\Omega=-L^{\textrm{on-shell}}$.

It starts with the definition of the stress-energy tensor:
\begin{align}
T_{\mu}^{\nu} &=- 2g^{\nu\rho}\frac{\partial \mathcal{L}^{m}}{\partial{g^{\mu \rho}}} + \delta_{\mu}^\nu \mathcal{L}^{m},\label{eq:Tmunu}\\ 
\mathcal{L}^{m} &= -\kappa \Tr\left[\frac{1}{2}\calF_{\mu\nu}\calF_{\rho\sigma}g^{\mu\rho}g^{\nu\sigma} + \calF_{\mu z}\calF_{\nu z}g^{\mu\nu}g^{zz}\right],
\end{align}
where the metric $g_{\mu\nu}$ is the full metric of the
Witten-Sakai-Sugimoto model.
We note that only the Yang-Mills part of the action appears, since the
CS term is independent of the metric.

From the stress-energy tensor we can extract the pressure, $P$, and
energy density $\mathcal{E}$ as
\beq
\mathcal{E} = -\int^{\infty}_{-\infty} \d z\sqrt{-g} T^0_0,\qquad
P=\frac{1}{3}\int^{\infty}_{-\infty} \d z \sqrt{-g}  T^i_i,
\eeq
which when used together with the holographic prescription
$\Omega=-L^{\textrm{on-shell}}$ should give the familiar thermodynamic
relations for homogeneous systems: 
\beq
\mathcal{E}=-P+\mu_B d_B, \qquad
PV=-\Omega.
\eeq

\subsection{Without \texorpdfstring{$S_{\rm CS}^\partial$}{S\_CS**del}}

As before, we start by considering the situation in which
$S_{\rm CS}^\partial=0$.
We can compute the pressure $P$ by dividing it into two contributions,
corresponding to the two terms in the stress-energy tensor
\eqref{eq:Tmunu}, $P=P^{(1)}+P^{(2)}$, with 
\beq
P^{(1)} = \frac{2\kappa}{3}\int^{\infty}_{-\infty}  \d z \Tr\left(h(z)F_{ij}^2+k(z)F_{iz}^2\right),\qquad
P^{(2)}=L_{\rm YM}=\int_{0}^{\infty} \d z\;\mathcal{L}_{\rm YM},
\eeq
where $L_{\rm YM}$ indicates the integrand of $S_{\rm YM}$ in
Eq.~\eqref{SYM} and the terms displayed in $P^{(1)}$ are the only
nonvanishing terms upon insertion of the homogeneous Ansatz
\eqref{eq:hom_ansatz}.

Since $P^{(2)}$ is trivial and already gives manifestly a part of the
Lagrangian density, we only need to compute $P^{(1)}$: Substituting
the homogeneous Ansatz, performing the traces and accounting for a
factor of two after trading the whole integration domain for only half
of the brane, we end up with 
\beq
P^{(1)}=2\kappa \int^{\infty}_{0}\d z\left[2h(z)H^4+k(z)H'^2\right],
\eeq
whose second term we can integrate by parts in order to make the equation of motion \eqref{eqH} manifest, leaving us with a boundary term as a result:
\beq
P^{(1)}=\frac{N_c}{8\pi^2}\int^{\infty}_{0}\d z\;\widehat{a}_0'H^3 + 2\kappa\left[\left(k(z)H'H\right)\right]_0^\infty.
\eeq
We can integrate by parts again to make the (bulk) CS term
\eqref{eq:CSbulk} appear,
at the price of obtaining a second boundary term:
\beq
P^{(1)}=\int_0^\infty \d z\;\mathcal{L}_{\rm CS}^{\rm bulk}- 2\kappa\left[\left(k(z)H'+\left.\frac{N_c}{16\pi^2\kappa}\widehat{a}_0H^2\right)H\right]\right\vert_{z=0}\label{POLD},
\eeq
where we made use of the fact that the boundary term only contributes
in the IR ($z=0$).
In the boundary term, we now recognize Eq.~\eqref{bdyeqHOLD}, so
that it vanishes on-shell at equilibrium (since $k(0)=1$).
In the end, we obtain the expected result 
\beq\label{PLAGOLD}
P = L_{\rm YM} + L_{\rm CS}^{\rm bulk} = L,
\eeq
and since from holography at $T=0$, we identify the on-shell
Lagrangian with the grand-canonical potential, we obtain $PV=-\Omega$.

The next quantity we need to compute is the energy density
$\mathcal{E}$. Again, we can divide the expression into two
contributions, corresponding to the two terms of Eq.~\eqref{eq:Tmunu}
and again the latter will trivially give minus the Yang-Mills
Lagrangian:
\beq
\mathcal{E}^{(2)}
= -\int_{-\infty}^{\infty} \d z\sqrt{-g}\,\mathcal{L}^{m}
= -L_{\rm YM},
\eeq
while the first term requires more attention as it will also produce
boundary terms in the effort to make the presence of $L_{\rm CS}$
manifest.
We start by computing the derivative with respect to $g^{00}$ to obtain:
\beq
\mathcal{E}^{(1)}=2\kappa \int_{0}^{\infty}\d z\;k(z)\widehat{F}_{0z}^2
=2\kappa \int_{0}^{\infty}\d z\;k(z)(\widehat{a}_0')^2,
\eeq
where the term after the first equality is the only nonvanishing one
upon insertion of the homogeneous Ansatz \eqref{eq:hom_ansatz}.
Again we can proceed to integrate by parts obtaining the kinetic term
of the equation of motion in exchange for a boundary term:
\beq
\mathcal{E}^{(1)}=
2\kappa\left[k(z)\widehat{a}_0\widehat{a}_0'\right]_{0}^{\infty}
-2\kappa \int_{0}^{\infty}\d z\;\partial_z(k(z)\widehat{a}_0')\widehat{a}_0.
\eeq
This time the only contribution from the boundary term comes from
$z=\infty$, as also noted in the hard-wall model in
Ref.~\cite{Bartolini:2022rkl}.
Making use of Eqs.~\eqref{eqa0} and \eqref{da0old}, the
bulk CS term and the chemical potential coupled to
the baryon density appear: 
\beq
\mathcal{E}^{(1)}= -L_{\rm CS}^{\rm bulk} + \frac{N_c}{2}\mu d.
\eeq
In the end, for the total energy density we obtain
\beq
\mathcal{E} = -L_{\rm YM}-L_{\rm CS}^{\rm bulk} + \frac{N_c}{2}\mu d.
\eeq
By making use of Eqs.~\eqref{PLAGOLD}, \eqref{mubold}, and
\eqref{dbold} we obtain the correct thermodynamic relation:
\beq
\mathcal{E}= -P+\mu_B d_B.
\eeq

\subsection{Including \texorpdfstring{$S_{\rm CS}^\partial$}{S\_CS**del}}

We now want to include the effects of the boundary term
$S_{\rm CS}^\partial$. Of course, the holographic dictionary still has
to be valid, and the system is still a homogeneous one, so consistency
requires us to still find the relation $PV=-\Omega = L$ on-shell at
equilibrium.

To obtain the result \eqref{POLD}, we only used the definition of
$T_{\mu\nu}$, which is sensitive only to the Yang-Mills action, and
the equations of motion, which are insensitive to boundary terms.
Hence the entire derivation is not altered by the presence of a
boundary term in the CS action and Eq.~\eqref{POLD} still
holds.
However, this time the boundary term in Eq.~\eqref{POLD} does not
vanish, but gives a contribution according to Eq.~\eqref{bdyeqHNEW}: 
\beq
P^{(1)}=\int_0^\infty \d z\;\mathcal{L}_{\rm CS}^{\rm bulk}
+\frac{3N_c}{32\pi^2}\left.\left[\widehat{a}_0 H^3\right]\right\vert_{z=0}
= L_{\rm CS}^{\rm bulk}+L_{\rm CS}^\partial\label{PNEW},
\eeq
so that in the end we again end up with the pressure equating the full
on-shell Lagrangian density at equilibrium, complying with the
holographic dictionary and the homogeneity of the system: 
\beq
P= L_{\rm YM}+L_{\rm CS}^{\rm bulk}+L_{\rm CS}^\partial = L,\qquad
PV = -\Omega.
\eeq

Let us now turn to the energy density: Here too the derivation from
the previous section holds up to the evaluation of the boundary terms,
since again only the Yang-Mills action and the equations of motion are
involved, so we only need to take a better look at 
\beq
\mathcal{E}^{(1)}=-2\kappa \int_{0}^{\infty}\d z\partial_z(k(z)\widehat{a}_0')\widehat{a}_0+2\kappa\left[k(z)\widehat{a}_0\widehat{a}_0'\right]_{0}^{\infty}.
\eeq
The integral still provides $L_{\rm CS}^{\rm bulk}$ upon insertion of
the equations of motion, but we note that $\widehat{a}_0'(z=0)=0$ no
longer holds; hence both boundaries will contribute now.
At both $z=\infty$ and $z=0$ we use of the explicit expression
\eqref{da0new} to obtain:
\beq
\mathcal{E}^{(1)}= -L_{\rm CS}^{\rm bulk}
+ \frac{N_c}{8}\mu d + \frac{3N_c}{32\pi^2} H^3(0)\widehat{a}_0(0),
\eeq
where we notice that the infrared contribution amounts exactly to
$-L_{\rm CS}^\partial$. The term proportional to $\mu d$ instead
provides again the correct quantity $\mu_B d_B$ once we make use of
the dictionary entries \eqref{dbnew} and \eqref{mubold}.
Combining $\mathcal{E}^{(1)}$ and $\mathcal{E}^{(2)}$, we obtain the
correct thermodynamic formula accounting for the presence of
$S_{\rm CS}^\partial$: 
\beq
\mathcal{E} = -L_{\rm YM}-L_{\rm CS}^{\rm bulk}-L_{\rm CS}^\partial + \frac{N_c}{8}\mu d = -P + \mu_B d_B.
\eeq

\section{Effects on observables}\label{sec:observables}

In this section, see how much impact the boundary term
$S_{\rm CS}^{\partial}$ has on a few selected observables relevant for
physics at finite densities; i.e.~we will compare the observables with
and without the presence of the boundary term, recalling that the
top-down model with the definition from string theory, should contain
this boundary term as part of the CS term.

\subsection{Saturation density}

Let us begin with evaluating the saturation density at the
phenomenological value (as derived from fitting the $\rho$ meson mass 
and the pion decay constant \cite{Sakai:2004cn}) of the 't Hooft coupling, 
i.e.~$\lambda=16.63$.
We simply vary the density $d$ and determine $\mu$ by the
thermodynamic equilibrium (i.e.~Eq.~\eqref{eq:mud_bulk} or
\eqref{eq:mud_full}), until we find the same value of the canonical
potential $\Omega$ for the baryon phase, as for the vacuum (which is
$\Omega=0$); this is the onset of the baryon phase and we define the
corresponding density as the nuclear saturation density, $d_0$.
We find
\begin{align}
  d_0^{\rm bulk} &= 0.436\left(\frac{M_{\rm KK}}{949\MeV}\right)^3 \fm^{-3},\\
  d_0^{{\rm bulk}+\p} &= 0.601\left(\frac{M_{\rm KK}}{949\MeV}\right)^3 \fm^{-3},
\end{align}
where $d_0^{\rm bulk}$ is computed with only the bulk CS
term, whereas $d_0^{{\rm bulk}+\p}$ is computed with the full
CS term. The mesonic fit of the model sets $M_{\rm KK}=949\MeV$, hence 
we can immediately see that the result closer to the phenomenological
$d_0^{\rm ph}=0.15\fm^{-3}$ is the one in which we neglect the
presence of $S_{\rm CS}^{\partial}$.
In order to obtain the nuclear saturation density of experiments
\cite{PhysRevC.102.044321}, the Kaluza-Klein scale would have to be
adjusted as:
\begin{align}
  d_0^{\rm bulk} &= 0.15\fm^{-3} \quad\Rightarrow\quad M_{\rm KK}=665.0\MeV,\non
  d_0^{{\rm bulk}+\p} &= 0.15\fm^{-3} \quad\Rightarrow\quad M_{\rm KK}=597.6\MeV,
  \label{eq:d0saturation}
\end{align}
for the bulk and full CS terms, respectively.

\subsection{Speed of sound}

We will now compute the speed of sound for the two cases, i.e.~with
and without the boundary term in the CS action taken into account,
which is given by
\beq
c_s^2 = \frac{d_B}{\mu_B}\frac{\p\mu_B}{\p d_B}
= \frac{d}{\mu}\frac{\p\mu}{\p d},
\eeq
and the only relation needed is $\mu(d)$ given by
Eq.~\eqref{eq:mud_bulk} and \eqref{eq:mud_full}, for the bulk CS and
the bulk+boundary CS term, respectively.
\begin{figure}[!htp]
  \begin{center}
  \includegraphics[width=0.6\linewidth]{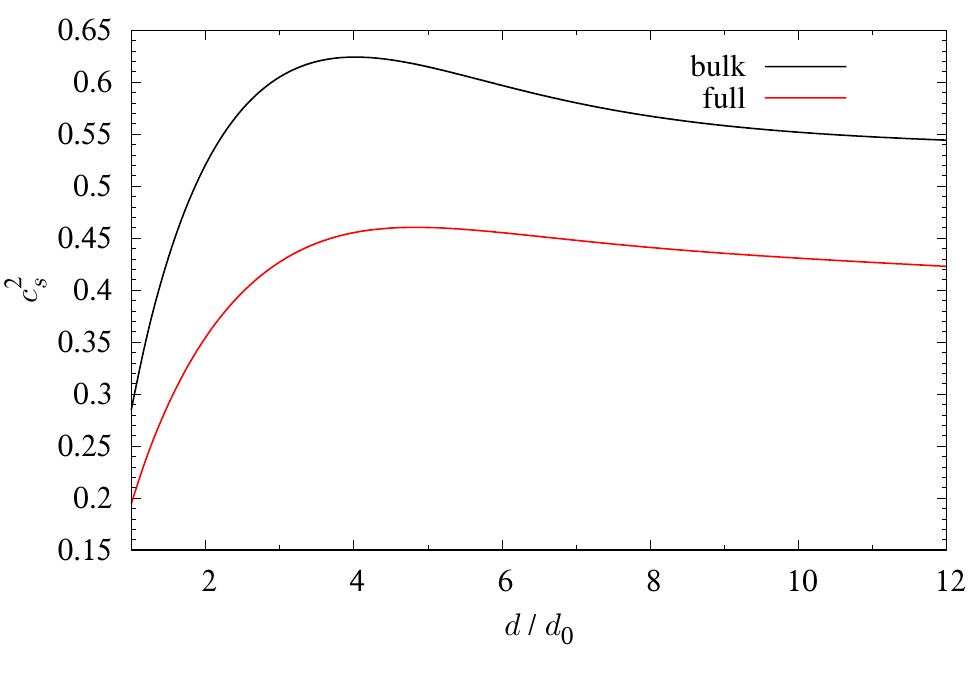}
  \caption{Sound speed squared for the two cases of the CS
    term without the boundary contribution (bulk) displayed with a
    solid black line and with it (full) displayed with a red solid
    line. The results are independent of $M_{\rm KK}$. }
  \label{fig:ss}
  \end{center}
\end{figure}
The result is shown in Fig.~\ref{fig:ss} for both cases.
Notice that the speed of sound does not depend on the calibration of
the KK scale.

\subsection{Equation of State}

Next, we will turn to the equation of state, which is a fundamental
relation for physics at finite density, with applications ranging from
heavy ion physics to neutron stars.
The equation of state (EOS) is a relation between the energy and the
pressure of a physical system and by the familiar thermodynamic
relations, $P=-\calE+\mu_B d_B$, where the chemical potential can
again be computed from Eq.~\eqref{eq:mud_bulk} and
\eqref{eq:mud_full}, for the bulk CS and the bulk+boundary CS term,
respectively.
For the thermodynamic relation, we have to recall the conversion
between the model parameters $\mu,d$ and the physical quantities
$\mu_B,d_B$, by Eqs.~\eqref{dbold} and \eqref{dbnew} for the density
for the bulk CS and the bulk+boundary CS term, respectively, and
Eq.~\eqref{mubold} for the chemical potential.

\begin{figure}[!htp]
  \centering
  \mbox{
  \subfloat[]{\includegraphics[width=0.49\linewidth]{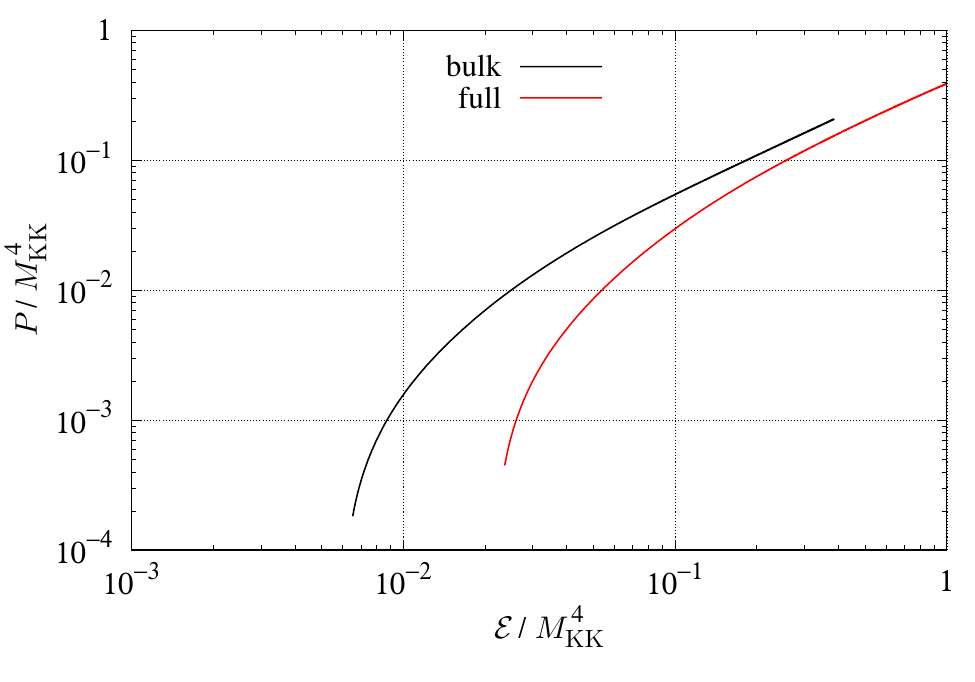}}
  \subfloat[]{\includegraphics[width=0.49\linewidth]{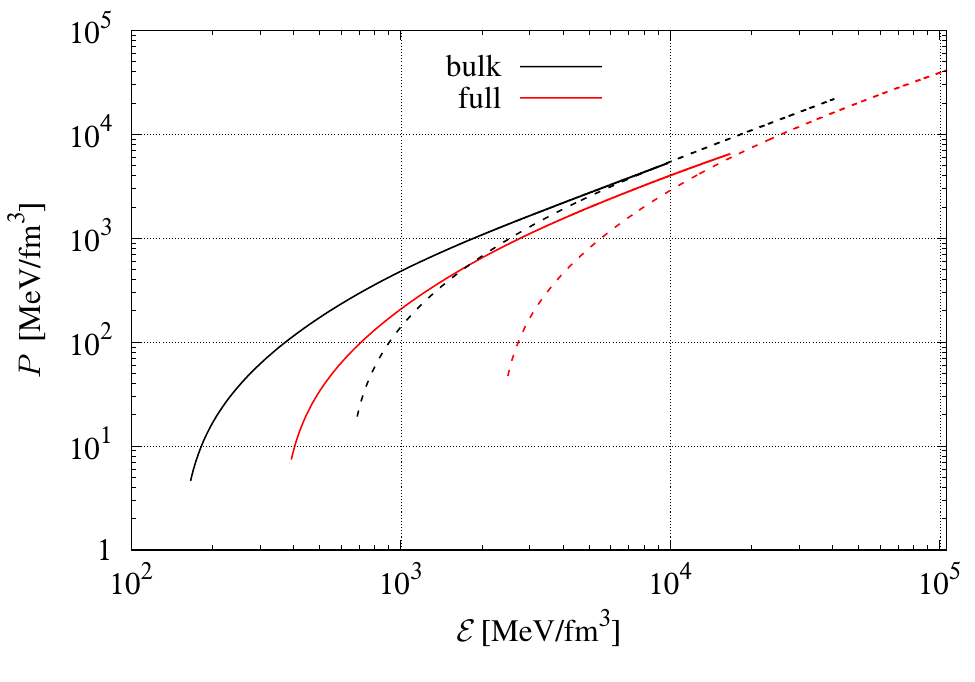}}}
  \caption{(a) The EOS in dimensionless units and (b) in calibrated
    units, for densities $d\in[1.1,20]d_0$ in the two cases of not
    including the boundary contribution in the CS term
    (black line) and with including it (red line).
    In (b) the solid lines correspond to the case that $M_{\rm KK}$ is calibrated after
    Eq.~\eqref{eq:d0saturation} such that the saturation density is
    physical, whereas the dashed lines correspond to the meson
    calibrated value of $M_{\rm KK}\sim949\MeV$ \cite{Sakai:2004cn}.
  }
  \label{fig:eos}
\end{figure}
The results for the EOS for symmetric nuclear matter are shown in
Fig.~\eqref{fig:eos}.
The panel (a) shows the dimensionless pressure and energy relation
(EOM) both normalized by the KK scale to the fourth power.
By using the calibration of having physical values of the saturation
density \eqref{eq:d0saturation}, the EOM is shown in physical units
with solid lines in panel (b) of the figure.
Using instead the mesonic fit of Ref.~\cite{Sakai:2004cn},
viz.~$M_{\rm KK}\sim949\MeV$, the EOM is shown with dashed lines.

\subsection{Symmetry Energy}

Next, we consider the nuclear symmetry energy, which measures the
energy cost in having more neutrons than protons, or vice versa.
For computing the symmetry energy, we follow
Ref.~\cite{Bartolini:2022gdf}, but we show the equivalence between
using a chemical isospin potential according to the traditional
holographic dictionary in App.~\ref{app:chempot}.
Rotating all gauge fields in isospin space $\calA\to a\calA a^\dag$
with a time-dependent SU(2) rotation, $a=a(t)$, and angular velocity
$\bchi=-\i\Tr(a^\dag\dot{a}\btau)$, we obtain the new time-dependent
Ansatz for the gauge fields:
\begin{align}
  \calA_0 = Ga\bchi\cdot\btau a^\dag + \tfrac12\wha_0,\qquad
  \calA_i = -\tfrac12(H a\tau^i a^\dag + L\chi^i),\qquad
  \calA_z = 0,
  \label{eq:symmetry_energy_Ansatz}
\end{align}
where the field $L$ has been turned on in order to satisfy the
equations of motion \cite{Bartolini:2022gdf}.
Notice that $\bchi\cdot\btau$ is the angular velocity, whereas
$a\bchi\cdot\btau a^\dag$ is the isospin angular velocity.
We want to study the theory without external fields turned on, while
the effects of isospin chemical potential have been already accounted 
for by the rotating Ansatz as shown in Ref.~\cite{Bartolini:2022gdf}, so the 
UV boundary conditions for the new fields, $G$ and $L$ are
\begin{align}
  G(\infty) &=0,\\
  L(\infty) &=0.
\end{align}
Computing the energy contribution arising from the new terms and the
isospin rotation, we find that they correspond to a kinetic term for
the Hamiltonian
\beq
H_{\rm kin} = \frac12V\Lambda\bchi\cdot\bchi
= \frac{I(I+1)}{2V\Lambda}, \qquad
\Lambda = \kappa\int\d z\left( 2h H^2(2G+1)^2 + k\left[(L')^2 + 4(G')^2\right]\right),
\label{eq:Hkin}
\eeq
where in the second equality, zeromode quantization has been
performed on the coordinates $a=a_0+\i a_i\tau^i$ on the 3-sphere,
yielding
$\pi_m=\frac{\p H}{\p\dot{a}_m}=4V\Lambda\dot{a}_m$,
$\pi_m^2=4I(I+1)$, with $I\in\frac12\mathbb{Z}_+$
being the isospin quantum number.
Using the relation between isospin and the difference between the
number of neutrons ($N$) and protons ($Z$)
\beq
2I = Z - N = -\beta B,
\eeq
with $B$ being the total number of nucleons, the symmetry energy can
readily be extracted from the Hamiltonian \eqref{eq:Hkin} as
\beq
S(d_B) = \frac{d_B}{8\Lambda},
\eeq
where $d_B$ is the physical nucleon density, given by
Eqs.~\eqref{dbold} and \eqref{dbnew} for the bulk CS and
the bulk+boundary CS term, respectively.

To consistently compute the IR boundary conditions we again follow the
method of vanishing IR boundary terms in the variation of the action,
the same as imposing thermodynamic equilibrium as previously shown.
The additional terms that appear in the Yang-Mills and CS
actions because of the rotation in SU(2) (ignoring second-order time
derivatives) are given by:
\begin{align}
S_\chi&=S_{\chi\rm YM} + S_{\chi\rm CS}^{\rm bulk} + S_{\chi\rm CS}^\partial
,\\
S_{\chi\rm YM}&= -\kappa\bchi^2\int \d^4x\int_0^{\infty}\d z\left[ k(L')^2  -
4k(G')^2 -8hH^2\left(G+\frac12\right)^2\right],\label{SchiYM}\\
S_{\chi\rm CS}^{\rm bulk}&= \frac{N_c}{4\pi^2}\bchi^2\int \d^4x\int_0^{\infty}
\d z\;\left[H^2 L G' + 2\left(G + \frac12\right) L H H'\right]
,\label{eq:CSchibulk}\\
S_{\chi\rm CS}^{\partial}&= \frac{3N_c}{16\pi^2}\bchi^2\int \d^4x\;
H^2(0)L(0)\left(G(0) + \frac49\right).
\end{align}
The equations of motion for all the fields do not depend on boundary
terms in the action, and they read
\begin{align}
h H^3 - \frac12\p_z(k H') - \frac{N_c}{32\pi^2\kappa}H^2\hat{a}_0' &= 0,\\
\p_z(k\hat{a}_0') + \frac{3N_c}{16\pi^2\kappa}H^2H' &= 0,\\
\p_z(k G') - 2h H^2\left(G + \frac12\right) + \frac{N_c}{32\pi^2\kappa} H^2L' &= 0,\\
\p_z(k L') + \frac{N_c}{8\pi^2\kappa} H\left[H G' + (1 + 2G)H'\right] &=0.\label{eq:eqL}
\end{align}

To determine the IR boundary conditions, we perform a variation of the
action and look at the IR boundary terms, imposing them to vanish. It is
now that we have to take into account whether $S_{\chi CS}^\partial$ is
present or not. As before, we consider first the situation in which we get
rid of it, so that the action at order $\bchi^2$ is given only by
$S_\chi =S_{\chi\rm YM} + S_{\chi\rm CS}^{\rm bulk} $. The IR boundary terms
that have to vanish are then given by
\begin{align}
\left[G'(0) + \frac{N_c}{32\pi^2\kappa}H^2(0)L(0)\right]\delta G(0)&=0,\label{eq:bdyeqGOLD}\\
L'(0)\dL (0)&=0\label{eq:bdyeqLOLD}.
\end{align}
The second equation of this set is trivially solved by imposing Neumann
boundary conditions for $L$, exactly as it happened for $\wha_0$.
Since we expect $L$ to have odd parity with respect to $z$, it will be
a discontinuous function if $L(0)\neq 0$.
If in turn $L(0)\neq 0$, $G'(0)\neq 0$ will be a nonvanishing
derivative at the IR tip and hence its derivative will not be
continuous due to the positive parity in $z$.

We can integrate the equation of motion \eqref{eq:eqL} once obtaining
\beq
\label{eq:integratedeqL}
k L'+\frac{N_c}{8\pi^2\kappa} H^2\left(G+\frac12\right)= {\rm const},
\eeq
where the right-hand side constant is a constant of motion, which can
be determined by using Eq.~\eqref{eq:bdyeqLOLD}, yielding
\beq
k L'+\frac{N_c}{8\pi^2\kappa} H^2\left(G+\frac12\right)
=\frac{N_c}{8\pi^2\kappa} H^2(0)\left(G(0)+\frac12\right),
\eeq
which when evaluated at $z\to\infty$ yields a nonvanishing axial U(1) 
current
\beq
\kappa[k L']_{z=\infty}
=\frac{N_c}{8\pi^2} H^2(0)\left(G(0)+\frac12\right).
\eeq

Let us now include the presence of $S_{\chi\rm CS}^{\partial}$: In this case
the IR boundary terms that have to vanish are:
\begin{align}
\left[G'(0)+\frac{N_c}{128\pi^2\kappa} H^2(0) L(0)\right]\delta G(0)&=0,
\label{eq:bdyeqGNEW}\\
\left[L'(0)+\frac{3N_c}{32\pi^2\kappa} H^2(0)\left(G(0)+\frac{4}{9}\right)
\right]\dL(0)&=0\label{eq:bdyeqLNEW}.
\end{align}
As can be seen, the IR boundary condition for $L$ is no longer a
Neumann condition.
Evaluating the constant of motion in the integrated equation of motion
for $L$, \eqref{eq:integratedeqL}, we obtain now
\beq
k L' + \frac{N_c}{8\pi^2\kappa}H^2\left(G+\frac12\right)
= \frac{N_c}{32\pi^2\kappa}H^2(0)\left(G(0)+\frac23\right),
\eeq
which when evaluated at $z\to\infty$ yields a different but still
nonvanishing axial U(1) current
\beq
\kappa[k L']_{z=\infty}
=\frac{N_c}{32\pi^2}H^2(0)\left(G(0)+\frac23\right).
\eeq

We note that both in the case of discarding the boundary term in
Eq.~\eqref{SCSfull} to arrive at the CS action \eqref{SCSbulk} and in
the case of keeping it, the axial U(1) current is turned on.
We have turned on isospin by isorotating the baryons and it is in fact
equivalent to using a chemical potential, see App.~\ref{app:chempot}.
We expect on general grounds that isorotation will induce a
nonvanishing U(1) current.
If, however, we would like to switch it off we can perform another
integration by parts at the level of $\bchi^2$ in the action, writing
the CS bulk and boundary terms at this order as
\begin{align}
S_{\chi\rm CS}^{\text{bulk,no-current}}&=-\frac{N_c}{4\pi^2}\bchi^2\int \d^4x\int_0^{\infty}
\d z\;H^2L'\left(G+\frac{1}{2}\right) ,\label{eq:CSchibulknocurrent}\\
S_{\chi\rm CS}^{\partial,\text{no-current}}&=\frac{N_c}{16\pi^2}\bchi^2\int \d^4x H^2(0)L(0)
\left(G(0)+\frac{2}{3}\right).
\end{align}
The integration by parts has been performed such that the field $L(z)$
is removed in favor of its derivative.
In this case, the IR boundary terms that have to vanish are then given
by
\begin{align}
G'(0)\delta G(0)&=0,\label{eq:bdyeqGOLDnocurrent}\\
\left[L'(0)+\frac{N_c}{8\pi^2\kappa} H^2(0)\left(G(0)+\frac{1}{2}\right)
\right]\delta L (0)&=0\label{eq:bdyeqLOLDnocurrent}.
\end{align}
The first equation of this set is trivially solved by imposing a Neumann 
boundary condition for $G$, exactly as it happened for $\wha_0$. 
We note that in this case the IR boundary equation for $L'(0)$ reduces
to the same form of the integrated equation of motion
\eqref{eq:integratedeqL}, with the constant of integration to be
determined by the boundary conditions. 
The condition \eqref{eq:bdyeqLOLDnocurrent} forces the constant to be
zero: As done with $\widehat{a}_0$ and the
associated charge, we can now use Eq.~\eqref{eq:integratedeqL} to
compute the current associated with $\widehat{A}_i$, (an axial current
since $\widehat{A}_i(z)$ is an odd function of $z$).
To do so, we evaluate Eq.~\eqref{eq:integratedeqL} at the UV boundary
to find
\beq
\kappa\left[kL'\right]_{z=\infty}= 0, 
\eeq
and since the left-hand side of the equation above is proportional to
the current, we conclude that we are working at zero axial U(1)
current. Note that by fixing the derivative $L'(0)$ we no longer have
the freedom of choosing the function $L(z)$ to be continuous at $z=0$:
The value $L(0)$ is now determined by the equations of motion and the
two boundary conditions, so $L(z)$, just like $H(z)$ is in general
discontinuous and odd.

We note that in all cases, the boundary conditions $L(0)=0$ and
$G'(0)=0$ are consistent with the variational principle, as the
Dirichlet boundary condition $L(0)=0$ eliminates the possibility of
the variation.
$G'(0)=0$ is also consistent in this case, since it is always
proportional to $L(0)$, which when vanishing implies a Neumann
condition for $G$ in the IR.
The boundary conditions found here, however, are expected to lower the
free energy slightly with respect to the simplistic, but consistent
boundary conditions $L(0)=G'(0)=0$.

In App.~\ref{app:chempot}, we have shown the result of
Ref.~\cite{Bartolini:2022gdf} adapted to the three cases of different
CS terms utilized in this section: The bulk CS, the full
(bulk+boundary) CS and the last ad-hoc construction that eliminates
the U(1) axial current.
The summary of the analysis in the appendix, is that only the bulk CS
term, $S_{\rm CS}^{\rm bulk}$, remains invariant under the specific gauge
transformation that connects the rotation of the isospin moduli to the
introduction of a finite isospin chemical potential as the UV boundary
value of the field $A_0^{a=3}$, as prescribed per usual in the
holographic dictionary.
We are then led to conclude that, also in this case, it is preferable
to neglect $S_{\rm CS}^{\partial}$, in order to preserve the equivalence
between angular velocity in isospin space and the isospin chemical
potential.

\begin{figure}[!htp]
  \centering
  \mbox{
  \subfloat[]{\includegraphics[width=0.49\linewidth]{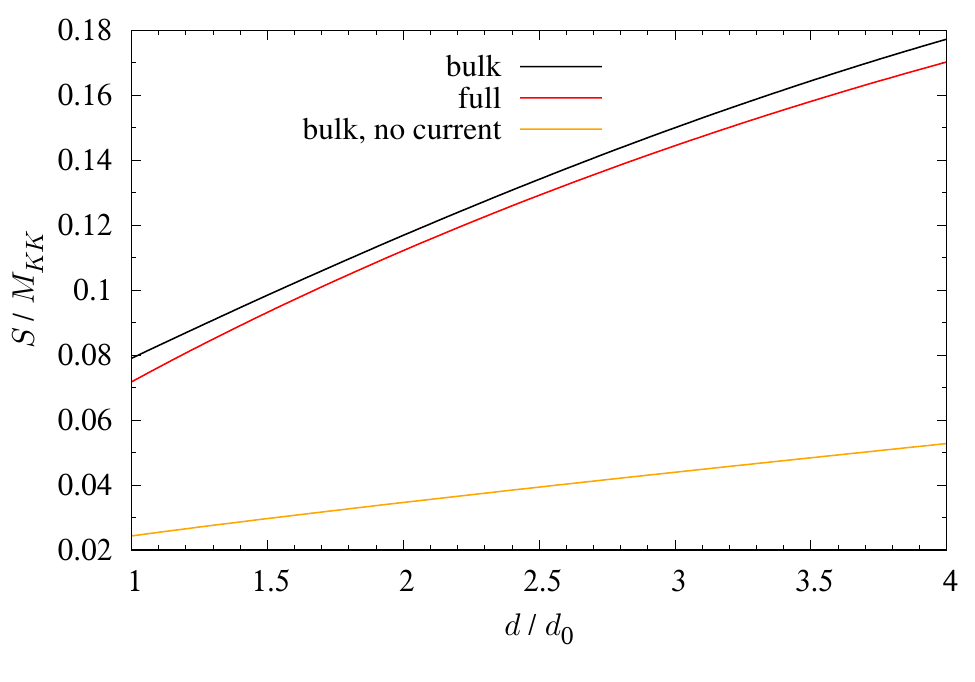}}
  \subfloat[]{\includegraphics[width=0.49\linewidth]{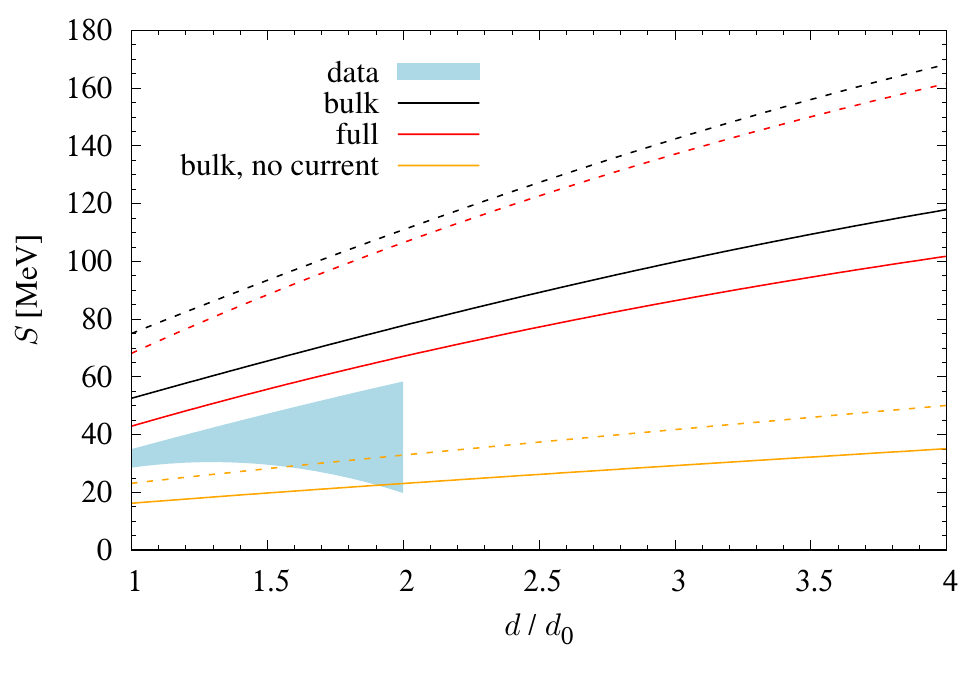}}}
  \caption{Symmetry energy in (a) dimensionless and (b) physical
    units, in the cases of not
    including the boundary contribution in the CS term
    (black line) and with including it (red line), as well as the
    choice of setting the external U(1) axial current to zero (orange line).
    In (b) the solid lines correspond to the case that $M_{\rm KK}$ is calibrated after
    Eq.~\eqref{eq:d0saturation} such that the saturation density is
    physical, whereas the dashed lines correspond to the meson
    calibrated value of $M_{\rm KK}\sim949\MeV$ \cite{Sakai:2004cn}.
    In (b) fitted data from the survey \cite{Li:2021thg} is shown with a
    light-blue shaded area.
    In this figure, $\lambda=16.63$.
  }
  \label{fig:esym}
\end{figure}
The results for the symmetry energy in the two cases of including the
boundary term in the CS action and not including it, are
shown in Fig.~\ref{fig:esym} with red and black colors, respectively.
Additionally, the ad-hoc choice of setting the U(1) axial current to
zero, according to the boundary conditions
\eqref{eq:bdyeqGOLDnocurrent}-\eqref{eq:bdyeqLOLDnocurrent} is shown
with orange curves.
In Fig.~\ref{fig:esym}(b) two calibrations are shown: viz.~that
corresponding to Eq.~\eqref{eq:d0saturation} such that the saturation
density is physical and that for which the standard rho meson
calibrated value of $M_{\rm KK}\sim949\MeV$ \cite{Sakai:2004cn}.
Finally, the phenomenologically expected region extracted from many
experiments via a fit \cite{Li:2021thg}, is shown with a light-blue
shaded area.

\subsection{Neutron Stars}

Our final observable to consider here, are the masses and radii of
neutron stars, ignoring fine details as the crust -- which however are
crucially important to obtain correct radii, that are approximately 1
or more kms larger than predicted by dense neutron matter at masses
around 1.4 solar masses -- and neglecting also isospin 
asymmetry, whose contribution in the gauge fields (the same we presented
in the previous section) is suppressed as $N_c^{-1}$.

The mass and radius of a single neutron star is obtain by solving the
Tolman-Oppen\-heimer-Volkoff (TOV) equations, which are given by
\begin{align}
  \frac{\d P}{\d r}&= -G(\calE+ P)\frac{m+4\pi r^3 P}{r(r-2Gm)},\\
  \frac{\d m}{\d r}&= 4\pi r^2 \calE,
\end{align}
where the nuclear physics input is in the form of the equation of
state, or more precisely, the inverse is needed: $\calE(P)$.
Due to the astronomical units in this system of equations, it will
prove useful to rescale the variables to dimensionless quantities, for
which the equations read
\begin{align}
  \frac{\d\widetilde{P}}{\d\tilde{r}}&=
  (\widetilde{\calE}+\widetilde{P})\frac{\tilde{m}+A\tilde{r}^3\widetilde{P}}{\tilde{r}(2\tilde{m} - \tilde{r}/B)},\\
  \frac{\d\tilde{m}}{\d\tilde{r}}&= A \tilde{r}^2\widetilde{\calE},
\end{align}
with the dimensionless conversion quantities
\begin{align}
  A &= \frac{4\pi r_0^2\calE_0}{m_0}
  \simeq 1.188911\left(\frac{M_{\rm KK}}{949\MeV}\right)^4, \\
  B &= \frac{Gm_0}{r_0}
  \simeq 1.477063,
\end{align}
where the physical and the dimensionless quantities are related as
$P=\calE_0\widetilde{P}$, $r=r_0\tilde{r}$,
$\calE=\calE_0\widetilde{\calE}$, and $m=m_0\tilde{m}$, and for
convenience, we have chosen $\calE_0=M_{\rm KK}^4$, $r_0=1\km$ and
$m_0=M_{\odot}$ (1 solar mass).
Finally, we denote by $R$ the radius $r(P)$ with $P=0$ and
correspondingly $M$ the mass $m(P)$ with $P=0$.

\begin{figure}[!htp]
  \centering
  \includegraphics[width=0.6\linewidth]{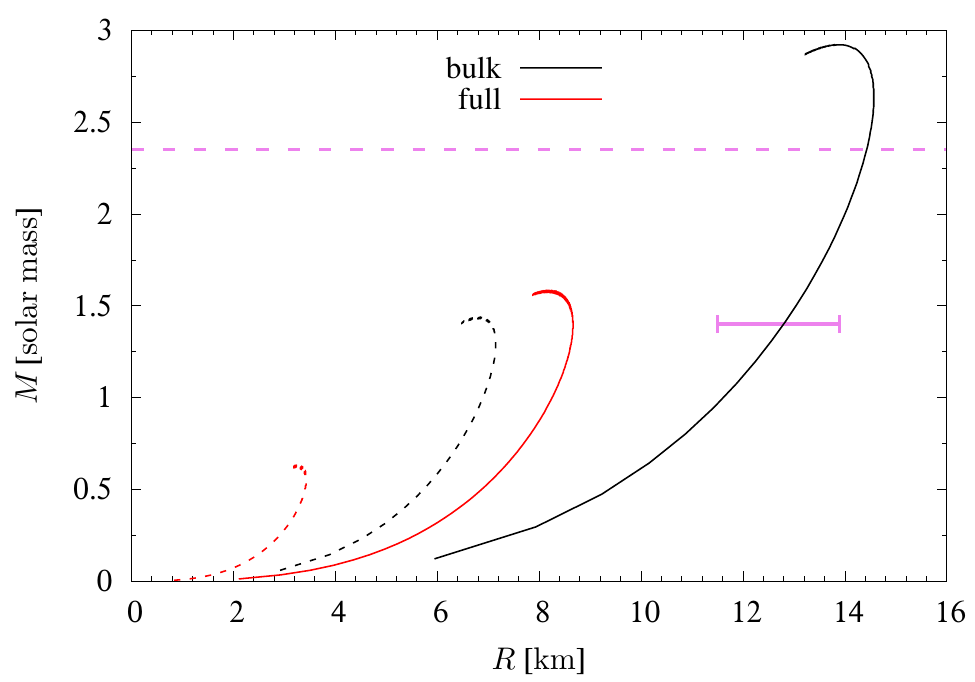}
  \caption{Neutron star mass and radius, in the two cases of not
    including the boundary contribution in the CS term
    (black line) and with including it (red line).
    For the solid lines, $M_{\rm KK}$ is calibrated after
    Eq.~\eqref{eq:d0saturation} such that the saturation density is
    physical, whereas the dashed lines correspond to the meson
    calibrated value of $M_{\rm KK}\sim949\MeV$ \cite{Sakai:2004cn}.
    The constraints on the radius at $1.4M_{\odot}$ come from
    J0740+6620, whereas the maximum observed stable mass at
    $2.35M_{\odot}$ is measured from PSR J0952-0607.
  }
  \label{fig:ns}
\end{figure}
In Fig.~\ref{fig:ns}, we show the results of the masses and radii for
the two cases of including the boundary term in the CS
action and not including it.
In the figure, two calibrations are shown: viz.~that corresponding to
Eq.~\eqref{eq:d0saturation} such that the saturation density is
physical and the standard meson calibrated value of
$M_{\rm KK}\sim949\MeV$ \cite{Sakai:2004cn}.
Since there is not taken any crust (softer matter at the surface of
the star) into account, the radius must be at least about 1km smaller
than the upper bound of the constraint from J0740+6620 (violet bar).
The maximum mass should be around $\sim2.35M_{\odot}$ but
probably smaller than $\sim2.5M_{\odot}$.
The sensitivity of the mass/radius curves to the 't Hooft coupling is
quite large, so even an order one change in the coupling from
$\lambda=16.63$, could make viable neutron star phenomenology. 
In Ref.~\cite{Bartolini:2023wis}, it was found that indeed it is
possible to fit the model to properties of nuclear matter at
saturation density, and obtain realistic neutron stars: There the
Dirichlet boundary condition $L(0)=0$ was used, so the problem of
choosing the correct CS term does not arise, as any boundary
term that we can choose is proportional to $L(0)$.
Imposing the Neumann boundary condition $L'(0)=0$ and neglecting
$S_{\rm CS}^\partial$ is expected to provide corrections to the
results obtained: Here we restricted ourselves to the analysis of the
general problem created by the presence of boundary terms, so we leave
the detailed computation of more realistic neutron stars to a future
work.

\section{Conclusion and discussion}\label{sec:conclusion}

In this paper, we have considered the subtle issue of a boundary term
in the Chern-Simons (CS) action, that is usually discarded in the
literature although the reason for discarding/ignoring it has not been
clear.
We propose a reason for discarding it, so that the baryon density
defined by the topological integral matches with the density that is
read off of the tails of the fields at the conformal boundary and
enters the thermodynamic relations of energy and pressure.
This method can straightforwardly be extended to other observable
quantities, at least in principle.
We have also shown that when the above matching holds, the CS action
is invariant under a gauge transformation that relates the method of
describing isospin via rotation of the moduli being the orientation in 
$\SU(2)\subset\SU(N_f)$, and that of introducing a chemical potential
as dictated by the standard holographic dictionary.

As a byproduct of our analysis, we find a very precise way of
computing the conditions for the thermodynamic equilibrium, without
even evaluating the action on field configurations.
We find that the thermodynamics relations can be made sense of both
with and without including the mentioned boundary term.
Nevertheless, for the above-mentioned formal reasons, we argue 
that that a single form of the CS action is preferred over
the others possible -- namely the one dubbed CS bulk.

We have shortly mentioned that the CS term is known in the WSS model
to be subtle in the literature already. Elaborating a little on this,
Hata and Murata made a proposal for changing the CS term in order to
reproduce a constraint on wave functions coming from a
Wess-Zumino-Witten term for $\SU(N_f)$ with $N_f>2$ \cite{Hata:2007tn},
but their proposal was pointed out by Lau and Sugimoto to be well
defined only on a compact 5-manifold and furthermore not being able to
reproduce the chiral anomaly of QCD \cite{Lau:2016dxk} (see our
App.~\ref{app:chiralanomaly} for an explanation on why the anomaly is
safe in our proposal).
The proposal of Lau and Sugimoto fixed this issue
with a more complicated expression for an alternative CS term.
A natural question would be whether the proposal of Lau and Sugimoto
would cancel the boundary term, that we are proposing to cancel in the
context of the homogeneous nuclear matter in the WSS model.
The answer is negative.
The proposal of Lau and Sugimoto is very simply put to split the
holographic direction up into two parts and add a total derivative
term that would correspond to what would come from a gauge
transformation of transforming the left-hand gauge fields into the
right-hand gauge fields, as well as an integral over a 5-cycle that
gives rise to the WZW term.
Since we consider only $\SU(2)$, the integral of the 5-cycle vanishes
and it is straightforward to show that the difference between our CS
terms at $z>0$ and $z<0$ is not a total derivative.
This means that there is no unitary gauge transformation that
connects the field configurations, which is not unexpected since the
fields are by construction made discontinuous. 

Finally, let us observe that the ambiguity arises from the lack of a
rigorous derivation of the homogeneous Ansatz itself: One can imagine 
that in the rigorous setup, where an infinite multi-instanton 
configuration is built, no IR boundary term would be generated due to the
smoothness of the fields, the extra dimensions of $\mathbb{R}^3$ would provide
the winding number, and the limit of very high density would not break this
topological feature.
When formulating the homogeneous Ansatz, the opposite is done: The 
fields are taken to be in the high density regime, where some kind of
``spatial average'' has been performed, and the instanton number is
then restored by hand by means of a discontinuity.
Despite intuitive arguments in favor of the presence of the
discontinuity \cite{Jarvinen:2021jbd}, no rigorous derivation with
true nonlinear solutions in the large density limit has been
performed.
We leave such laborious task for future work.

\subsection*{Acknowledgments}

We thank Matti J\"arvinen for discussions.
The work of L.~B.~is supported by the National Natural Science
Foundation of China (Grant No.~12150410316). 
S.~B.~G.~thanks the Outstanding Talent Program of Henan University and
the Ministry of Education of Henan Province for partial support.
The work of S.~B.~G.~is supported by the National Natural Science
Foundation of China (Grants No.~11675223 and No.~12071111) and by the
Ministry of Science and Technology of China (Grant No.~G2022026021L).

\appendix

\section{Equivalence between \texorpdfstring{$\SU(2)$}{SU(2)}
  isospin rotation and external chemical potential}\label{app:chempot}

We start with the Ansatz \eqref{eq:symmetry_energy_Ansatz} and perform
a time-dependent gauge transformation in $\SU(2)$:
\beq
\calA \to b\calA b^\dag - \i b\d b^\dag,\qquad b=b(t)\in\SU(2),
\eeq
and since we want to eliminate the rotation matrices $a$ from the
Ansatz, we choose $b=a^{-1}$.
Using that $-\i a^\dag\dot{a}=\frac12\bchi\cdot\btau$, we obtain the
gauge fields
\beq
\calA_0 = \left(G+\frac12\right)\bchi\cdot\btau + \frac12\wha_0, \qquad
\calA_i = -\frac12\left(H\tau^i+L\chi^i\right), \qquad
\calA_z = 0.
\eeq
The function $G$ still has the UV boundary condition $G(\infty)=0$,
but a simple change of variables is convenient
\beq
\widetilde{G}=G+\frac12, \qquad \widetilde{G}(\infty)=\frac12.
\eeq
which leads to the theory with an isospin chemical potential turned
on:
\beq
\calA_0(\infty) = \frac12\bchi\cdot\btau = \frac12\mu_I\tau^3,
\eeq
where in the last equality, we have identified the isospin
angular velocity as $\bchi=(0,0,\mu_I)$. The $\tau^3$ direction of a
spin (or isospin) system is conventional.

Since the difference between using the isospin
angular rotation and the isospin chemical potential dictated by the
holographic dictionary is merely a gauge transformation, this changes
nothing for the Yang-Mills part of the action nor for the equations of
motion -- they remain unchanged.

Now the CS term is not gauge invariant and therefore the
different versions of the CS terms will change differently
when performing this gauge transformation.
Ideally we want the CS term not to change under this gauge
transformation so as to keep the free energy and boundary conditions
of the system invariant.

We start with the case of the bulk CS term \eqref{eq:CSchibulk} and
find by explicit computations that it does not change under the
above-described gauge transformation.
This can readily be inferred from Eq.~\eqref{eq:CSbulk}, where we can
see that the trace of $F_{\alpha_2\alpha_3}F_{\alpha_4\alpha_5}$ is
invariant since the non-Abelian field strength transforms covariantly
as $F\to bFb^\dag=a^\dag Fa$, the Abelian field strength $\widehat{F}$
does not transform, and the only dependence on $\calA_\mu$ is on the
Abelian part, which is untouched by an $\SU(2)$ gauge transformation.
Hence, the action is unchanged, the equations of motion are unchanged
and the IR boundary conditions remain exactly those of
Eqs.~\eqref{eq:bdyeqGOLD}-\eqref{eq:bdyeqLOLD}.

Moving to the case of including the boundary term in the CS action,
i.e.~using the full CS term, it is clear from Eq.~\eqref{eq:CSbulk}
that $A$ dependence is unavoidable.
An explicit computation reveals that the full CS term changes by
\beq
\frac{N_c}{96\pi^2}\int\d^4x\int_0^\infty\d z\;\p_z(LH^2)\bchi^2,
\eeq
which in turn changes the IR boundary condition for $L$ from
Eq.~\eqref{eq:bdyeqLNEW} to
\beq
\left[L'(0)
  + \frac{3N_c}{32\pi^2\kappa}H^2(0)\left(G(0) + \frac12\right)\right]\dL(0)=0.
\eeq

Since the bulk CS term is invariant under the gauge transformation
between the isospin rotation and the isospin chemical potential
interpretation of the theory, it is in that sense preferred compared
to the full CS term.

Finally, let us note that the CS term \eqref{eq:CSchibulknocurrent} is
also not invariant under the gauge transformation that is necessary
for switching from the isospin rotation to the isospin
chemical potential realization of the theory.
This can be seen from writing down the CS term, in the gauge
$\calA_z=0$, in the form
\beq
\frac{N_c}{32\pi^2}\epsilon^{MNKL}\left[
  \Tr(\calA_0\calF_{MN}\calF_{KL}) - \widehat{A}_0\Tr(\calF_{MN}\calF_{KL})\right].
\eeq
Clearly, the first term changes under the gauge transformation as it
contains the non-Abelian gauge potential.
We find that it gives rise to the CS action
\eqref{eq:CSchibulknocurrent} for the gauge transformed fields, but
changes to
\beq
-\frac{N_c}{4\pi^2}\bchi^2\int\d^4x\int_0^\infty\d z\;H^2L'G,
\eeq
when transforming back to the original gauge and hence differs from
Eq.~\eqref{eq:CSchibulknocurrent}.

\section{QCD anomaly and Chern-Simons forms}
	\label{app:chiralanomaly}
Here we want to show that each of the CS terms we presented in this work 
reproduces the QCD global anomaly: The discussion follows  
App.~C of Ref.~\cite{Domenech:2010aq} closely, but specialized to our
notation.
Since it can be thought that correctly reproducing the anomaly could
be used as a criterion for determining the ``physical'' CS
term, it is useful to show that this is not the case.

First of all, we start by recalling the full CS term:
\beq\label{eq:SCS}
S_{\rm CS}=\frac{N_c}{24\pi^2}\int_{5D}\omega_5(\calA).
\eeq
with $\omega_5$ being the standard CS five-form given by
\beq
\omega_5=\Tr\left[3\widehat{A}\wedge F^2+\widehat{A}\wedge\widehat{F}^2+\d\left(\widehat{A}\wedge\left(2F\wedge A-\frac{\i}{2}A^3\right)\right)\right], \quad \omega_5^{\rm SU(2)}=0.
\eeq
A variation of \eqref{eq:SCS} with gauge function $\alpha(z)$ whose
boundary values reduce to the parameters of a chiral transformation
$(\alpha_L,\alpha_R)$ is given by the formula:
\beq\label{eq:varCSfull}
\delta_{\alpha} S_{\rm CS}=  \frac{N_c}{24\pi^2}\int d^4x\left\{ \left[ \omega_4(\alpha(z), A ) \right]^{+\infty}_{0^+} -  \left[ \omega_4(\alpha(z), A ) \right]^{-\infty}_{0^-}\right\},
\eeq
where we introduced the four-form 
\beq
\omega_4(\alpha(z),A(z))= \Tr\left[ \alpha \d \left( A\d A -\frac{\i}{2} A^3 \right) \right].
\eeq
We can rearrange the terms into UV and IR contributions, keeping in
mind that the UV values of the gauge fields are holographically mapped
to the sources for the four-dimensional theory as $A(+\infty) = l$,
$A(-\infty )= r$.
The resulting expression for the variation of the CS term is:
\beq
\delta_{\alpha} S_{\rm CS}=  \frac{N_c}{24\pi^2}\int d^4x\left\{ \left[ \omega_4(\alpha_L, l ) - \omega_4(\alpha_R, r ) \right] -  \left[ \omega_4(\alpha(z), A(z) )\right]^{z=0^+}_{z=0^-}\right\}.
\eeq
The vanishing of the $z=0$ term together with the IR variation of the
Yang-Mills action will determine the IR boundary conditions as we have
explained, while the UV boundary term correctly reproduces the QCD
anomaly in its symmetric form.
  
We can now divide the standard five-form $\omega_5$ in two terms as
before, a bulk and a boundary term $\omega_5=\overline{\omega}_5+\d X$.
This separation is completely arbitrary, but to illustrate the
procedure we will use the choice we employed throughout the paper:
\beq
\overline{\omega}_5 && \equiv \Tr\left[3\widehat{A}\wedge F^2+\widehat{A}\wedge\widehat{F}^2\right],\\
\d X && \equiv  \d\Tr\left[\left(\widehat{A}\wedge\left(2F\wedge A-\frac{\i}{2}A^3\right)\right)\right].
\eeq
If we discard the boundary term $\d X$, we are choosing a nonstandard
form of the CS term: If the fields were continuous this
would not affect any physics, but for our discontinuous fields, we
argued that this choice has observable consequences.
However, as far as the anomaly is concerned, we can see that
everything proceeds in the same fashion: We can perform a variation of
$S_{\rm CS}^{\rm bulk}$ to obtain  
\beq
\delta_{\alpha} S_{\rm CS}^{\rm bulk} =  \frac{N_c}{24\pi^2}\int d^4x \left\{\left[\overline{\omega}_4(\alpha_L, l )-\overline{\omega}_4(\alpha_R, r ) \right]-\left[\overline{\omega}_4(\alpha,A)\right]^{z=0^+}_{z=0^-}\right\},
\eeq 
where we have introduced $\overline{\omega}_4$ that reads
\beq
\overline{\omega}_4(\alpha,A)= \Tr\left[3\widehat{\alpha}\wedge F^2 + \widehat{\alpha}\wedge\hF^2\right].
\eeq
Once again, the IR terms will determine the boundary conditions, which
will be different than the ones determined before (hence all the
different physics), while the UV term reproduces again the anomaly in
another form: To see that the two forms are equivalent, is sufficient
to note that they only differ by the addition of a local counter term, 
the variation $\delta_{\alpha}X$ of the four-form $X$.
The difference between the two forms of the anomaly is then given by
\beq
\frac{N_c}{24\pi^2}\int d^4x \left\{ \left[ \omega_4(\alpha_L, l ) - \omega_4(\alpha_R, r ) \right] -\left[\overline{\omega}_4(\alpha_L, l )-\overline{\omega}_4(\alpha_R, r ) \right]\right\} = \non
\frac{N_c}{24\pi^2} \int d^4x \left[  \delta_{\alpha_L}X(l)-\delta_{\alpha_R}X(r)\right].
\eeq 
Since $X$ only depends on the sources $l,r$, which are taken to have
vanishing physical value, the form in which we cast the anomaly has no
consequences.

\bibliographystyle{JHEP}
\bibliography{bib.bib}
\end{document}